\documentclass[superscriptaddress,aps,prl,twocolumn,preprintnumbers,amsmath,amssymb,floatfix,showpacs]{revtex4-1}

\usepackage{graphicx}
\usepackage{dcolumn}
\usepackage{bm}
\usepackage{amssymb}
\usepackage{xr}

\newcommand{\ket}[1]{\ensuremath{\left|#1\right\rangle}}
\newcommand{\braket}[2]{\left\langle #1\middle|#2\right\rangle}

\begin{document}

\title{Proposal for a loophole-free violation of Bell inequalities \\ with a set of single photons and homodyne measurements}

\author{Jean Etesse}

\affiliation{Laboratoire Charles Fabry, Institut d'Optique, CNRS, Universit\'e Paris Sud \\ 2 avenue Augustin Fresnel, 91127 Palaiseau cedex, France}

\author{R\'emi Blandino}

\affiliation{Laboratoire Charles Fabry, Institut d'Optique, CNRS, Universit\'e Paris Sud \\ 2 avenue Augustin Fresnel, 91127 Palaiseau cedex, France}

\author{Bhaskar Kanseri}

\affiliation{Laboratoire Charles Fabry, Institut d'Optique, CNRS, Universit\'e Paris Sud \\ 2 avenue Augustin Fresnel, 91127 Palaiseau cedex, France}

\author{Rosa Tualle-Brouri}

\affiliation{Laboratoire Charles Fabry, Institut d'Optique, CNRS, Universit\'e Paris Sud \\ 2 avenue Augustin Fresnel, 91127 Palaiseau cedex, France}
\affiliation{Institut Universitaire de France, 103 boulevard Saint-Michel, 75005 Paris, France}

\date{\today}


\begin{abstract}
We demonstrate that different kind of mesoscopic quantum states of light can be efficiently generated from a simple iterative scheme using homodyne heralding.
These states exhibit strong non-classical features, and are of great interest for many applications such as quantum error-correcting codes or fundamental testings.
On this basis we propose a protocol allowing a large loophole-free violation of a CHSH-type Bell inequality
with a remarkable robustness to line losses.
\end{abstract}


\maketitle

The violation of Bell inequalities is a crucial test for the foundations of quantum mechanics, to rule out classical mechanisms as the origin of quantum correlations \cite{Bell}. But up to now all experimental demonstrations of such violation left opened two so-called ``loopholes": the ``locality" loophole, which arises when the separation between the measured states is not large enough to completely discard the exchange of subluminal signals during the measurements \cite{Aspect,Weihs}; and the ``detection-efficiency" loophole, which occurs when the particle detectors are not efficient enough to warrant that the detected events are representative of the whole ensemble \cite{Pearle,Garg}.

The detection efficiency loophole has been closed with information encoded on atoms and ions \cite{Rowe,Pironio}, but one can hardly imagine separating such systems far enough to close the locality loophole. Light is a better candidate to close this second loophole as it can easily cover long distances \cite{Aspect,Weihs}, but it is difficult to detect photons with a high efficiency. The minimal efficiency required to close the detection efficiency loophole with the CHSH inequality \cite{Clauser} is $82.8\%$ in the symmetric case, while it is $66.6\%$ with the Eberhards inequality \cite{Eberhard}. Such a value was recently reached \cite{Giustina}, closing the loophole with global detection efficiencies between $70\%$ and $80\%$, using transition edge sensors \cite{Lita}. These cryogenic sensors are however not really widespread technologies and that is the reason why both loopholes are still not closed together. But even if this goal could be achieved in the near future, a loophole free violation of Bell inequalities is not only a major test for quantum mechanics: it can meet concrete applications like device independent quantum key distribution \cite{Pironio09}, and for that reason it can be interesting to obtain such a violation with affordable devices.

Homodyne measurements could be an interesting alternative as they are not subject to the detection efficiency loophole: every measurement will lead to a result. Different schemes have been proposed to obtain a violation with homodyne measurements, but they require either measurements \cite{Banaszek,Stobinska} or states \cite{Munro,Wenger,Cavalcanti,Acin} that seem practically unfeasible. Some proposals are based on more realistic setups \cite{Nha,Garcia} but provide only a small hardly measurable violation. Some hybrid schemes, mixing homodyne measurements with photon detection, have been recently proposed \cite{Brask12}, allowing a detection efficiency threshold of about $65\%$ with feasible states, but quite more complex states have to be considered in order to obtain a lower value \cite{Quintino}.

\begin{figure}[!h]
\begin{center}
\includegraphics[width=8.6cm]{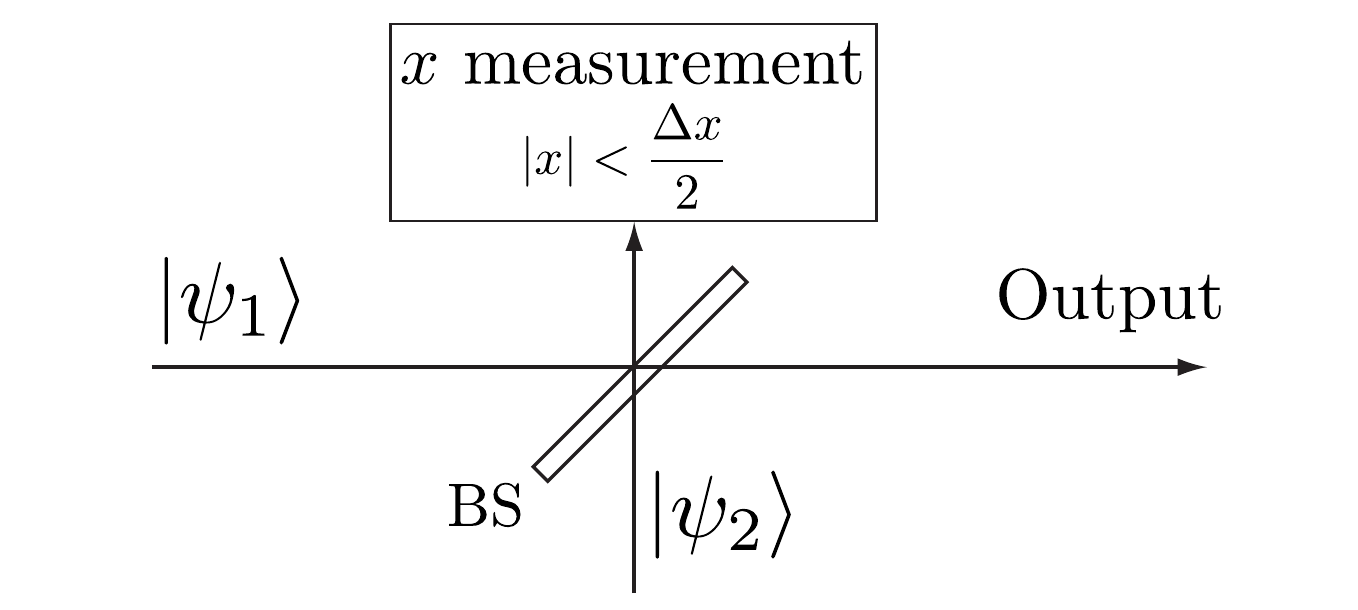}
\caption{{Elementary stage involved in the present protocol: two states are mixed on a symmetric beamsplitter, and the desired output state is generated owing to a heralding event based on a quadrature measurement.}} \label{fig1}
\end{center}
\end{figure}

\begin{figure}[!h]
\begin{center}
\includegraphics[width=8.6cm]{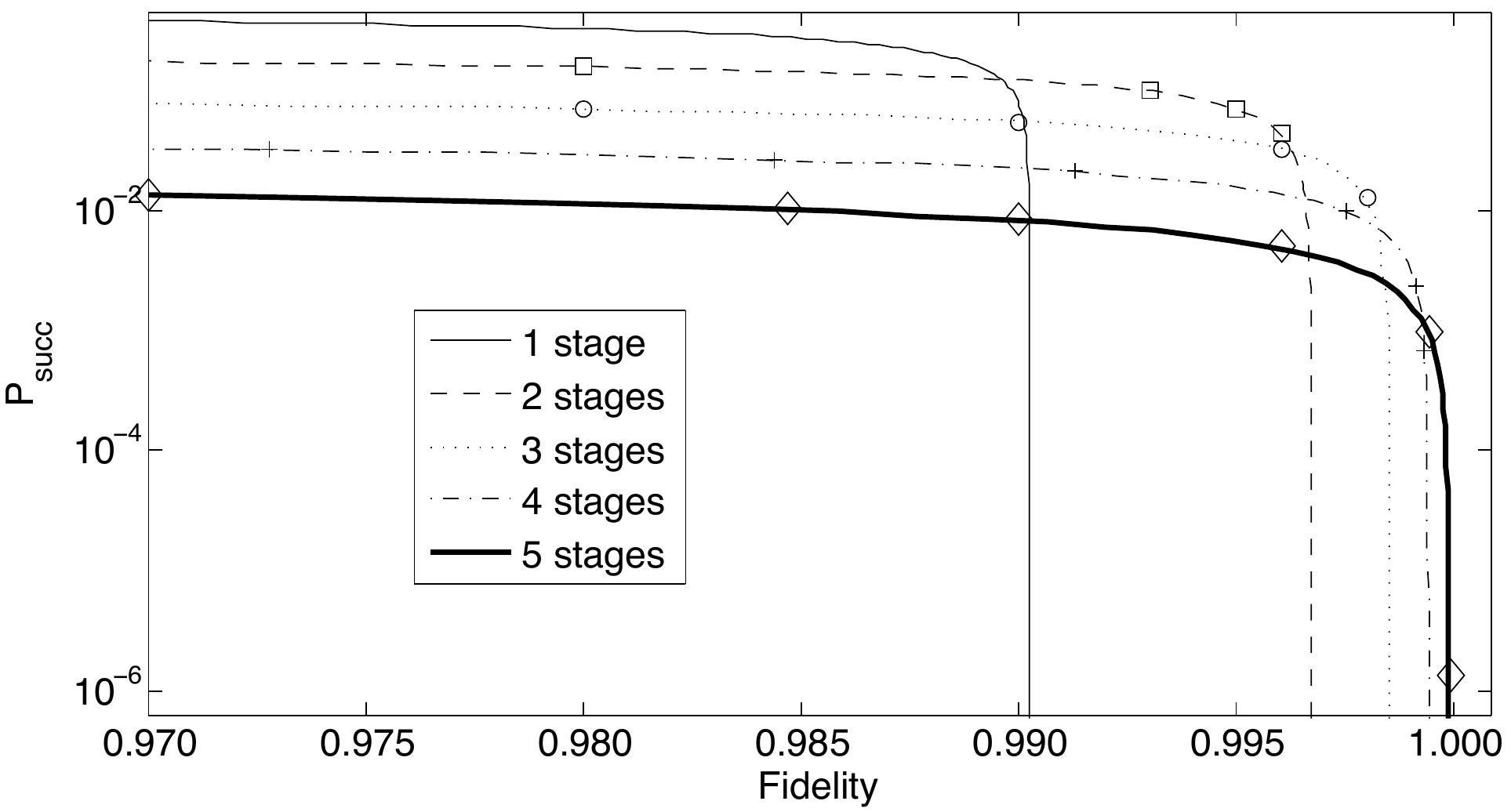}
\caption{{Success probability as a function of the fidelity of the state obtained after $p$ iterations of the breeding scheme of Fig.\ref{fig1} with the nearest squeezed SCS, for $p$ varying from $1$ to $5$. The lines are obtained with $\Delta x_{l+1} = 1.3 \Delta x_l$, while the different symbols correspond to optimized $\Delta x_l$.}} \label{fig2}
\end{center}
\end{figure}

In this paper we present a realistic protocol to implement the method proposed in [16], which allows a maximal violation of the CHSH inequality with simple homodyne measurements performed on a quite complex state. We show that this state can be efficiently generated using a method based on an iterative application
of the setup depicted in Fig. \ref{fig1}:  two states
$\psi_1(x)$ and $\psi_2(x)$ (where $x$ is a quadrature) are mixed on a 50:50 beamsplitter, and owing to the heralding event $|x|<\Delta x/2$ one implements, if $\Delta x$ is small enough, the non-linear operation:
\begin{equation}
(\psi_1(x),\psi_2(x))\ \ \tilde{\longrightarrow}\ \  \psi_1(\frac{x}{\sqrt{2}})\psi_2(\frac{x}{\sqrt{2}}).
\label{breedingbase}
\end{equation}
In the case where the two input states are identical, such a setup allows cat breeding operation \cite{Lund}, which can thus be performed with quadrature measurements \cite{Takeoka,Brask}. This can be understood as follows: if this protocol is iterated $p$ times, it will output the final state (with $n=2^p$):

\begin{equation}\label{Breeding}
\psi_{\rm out}(x)=N \psi^n(x/\sqrt{n})
\end{equation}

\noindent where $N$ is a normalization factor. If for instance the first input states are pure single photons, with $\psi(x)\propto x\exp(-x^2/2)$, the resulting state will be arbitrarily close (at large $n$) to an even Schr{\"o}dinger cat state (SCS) $\ket{\alpha}+\ket{-\alpha}$  of amplitude $\alpha=\sqrt{n}$ and with 3dB squeezing \cite{OurjoumtsevCat}. SCS can therefore be generated from single photon states with an iterative homodyne heralding and, as was also noticed in \cite{Laghaout}, this can be performed in a really efficient way as good fidelities can be obtained with quite large heralding width: we have searched the optimal width $\Delta x_l$ at each stage $l$ ($1\leq l\leq p$) in order to maximize the success probability for a given fidelity with the nearest SCS. We empirically found that the choice $\Delta x_{l+1} = 1.3 \Delta x_l$ is very close to the best one as shown in Fig. \ref{fig2}.\\
 In the following, we will define the success probability $P_{\rm succ}$ as the ratio of the minimal number of resources needed for the protocol (single photon states) to the average number of resources needed for a success event.
 Thanks to the iterative nature of the protocol, which allows to store intermediate states in quantum memories, this
 success probability (pictured in Fig. 2) does not decrease much faster than 1/n ($\sim1.3/n^{1.3}$ for instance at $\mathcal{F}=97\%$), with an average number of resources of the order of $n^2$.
 This last quantity can be large, but should be manageable in the near future with time and space multiplexing.

Breeding with any number of photons $n$ can also be performed by writing the binary expression of $n$. For instance, if $n=7=111_2$ the breeding will consist in the mixing of a 0-stage breeded cat ($2^0$, a single photon) with a 1-stage breeded cat ($2^1$) to form a state $\psi_{c,1}$, which will then be mixed with a 2-stages breeded cat ($2^2$) to form the output cat. The parity of this cat will be the same as the parity of $n$.\\

The non-linear transform in Eq.(\ref{breedingbase}) can find applications beyond SCS generation. To see this, it is interesting to understand how it can transform a single photon state into a SCS. The wavefunction $\psi(x)$ of the pure single photon presents two extrema at $x=\pm1$, and can be approximated at the vicinity of these extrema as $\psi(x)\approx \psi(\pm1)[1-(x\mp1)^2]\approx \psi(\pm1)\exp[-(x\mp1)^2]$. By raising this last term to the $n^{th}$ power, it results two thin Gaussians of width $1/\sqrt{n}$ (that are rescaled in Eq.(\ref{Breeding}) by the $1/\sqrt{n}$ term). Out of the vicinity of these extrema, the output function is smaller than the tails of the Gaussians and is therefore negligible.\\

 The same reasoning can be extended to any other wavefunction with local extrema. With a sine function for instance, applying Eq.(\ref{Breeding}) will result in a comb of Gaussian peaks. Such a sine function can be obtained by simply applying a phase shift $\pi/2$ to a squeezed cat state (amplitude $\alpha$, squeezing $s'$), which gives (with a normalization factor here only valid at high alpha):
\begin{equation}\label{SCSeven}
\psi^{s'}_+(x)=2(\frac{s'^2}{\pi})^{1/4}\exp(-s'^2\frac{x^2}{2})\cos(s'\sqrt{2}\alpha x)
\end{equation}
for an even squeezed cat state, and
\begin{equation}\label{SCSodd}
\psi^{s'}_-(x)=2(\frac{s'^2}{\pi})^{1/4}\exp(-s'^2\frac{x^2}{2})\sin(s'\sqrt{2}\alpha x)
\end{equation}
for an odd squeezed cat state.

Comb states were first introduced by D. Gottesman and coworkers in \cite{Gottesman} as a mean to construct error-correcting codes embedding a finite dimensional space in a system described by continuous quantum variables. Here we propose a simple method to encode such states on a travelling optical mode using only linear optics and homodyne detection. Let us first  define these states explicitly: if $G_s(x)\equiv(\pi s^2)^{-1/4}\exp[-x^2/(2s^2)]$ we introduce, following \cite{Gottesman}, the comb states

\begin{eqnarray}\label{comb_def}
\braket{x}{\bar 0}_a^{s,s'}&=&\sqrt{a}G_{1/s'}(x)\sum_k G_s(x-ka)\\
\braket{x}{\bar 1}_a^{s,s'}&=&\sqrt{a}G_{1/s'}(x)\sum_k G_s\Big(x-(k+\frac{1}{2})a\Big),
\end{eqnarray}

\noindent where the normalization factor $\sqrt{a}$ is valid only when $s\ll a\ll 1/s'$, which we will call the comb state condition: the Gaussian peaks have to be thinner than their separation, and a significant number of peaks is required within the overall Gaussian envelope.\\
We propose a ``comb breeding" protocol, by feeding the setup of Fig. \ref{fig1} with \textit{e.g.} the cat state (3) and by iterating the protocol $p'$ times.
It is clear from the previous considerations that such a protocol, whose action is summarized in Eq.(2), will generate a good approximation of $\ket{\bar{0}}^{s,s'}_a$, with $a=2^{p'/2}a_0$, $a_0=\pi/(\sqrt{2}s'\alpha)$ and $s=a_0/\pi$. At each iteration step the spacing $a$ is multiplied by $\sqrt{2}$.

This point can be understood in another way by computing the action of a symmetric beamsplitter (BS) on comb states. For instance one has:
\begin{equation}\label{Bell}
\ket{\bar 0,\bar 0}_{a}^{s,s'}\rightarrow \frac{1}{\sqrt{2}}\Big[\ket{\bar 0,\bar 0}^{s,s'}_{a\sqrt{2}}+\ket{\bar 1,\bar 1}^{s,s'}_{a\sqrt{2}}\Big]
\end{equation}
\noindent which is one of the Bell states in the comb states basis. The comb breeding protocol can then be easily understood: the measurement $x\simeq 0$ selects the comb state $\ket{\bar 0}^{s,s'}_{a\sqrt{2}}$ in the heralding port of the beamsplitter, and a comb state $\ket{\bar 0}^{s,s'}_{a\sqrt{2}}$ is therefore generated. The main point here concerns the fact that the detection of $\ket{\bar 0}^{s,s'}_{a\sqrt{2}}$ is not limited to $x\simeq0$: any event like
\begin{equation}\label{HeraldingCond}
\forall k\in\mathbb{Z},\qquad|x-ka\sqrt{2}|<\Delta x/2
\end{equation}
can be considered as successful, what will considerably increase the success rate of the heralding process.\\
\noindent Similarly, any event like
\begin{equation}\label{OneHeraldingCond}
\forall k\in\mathbb{Z},\qquad|x-(k+\frac{1}{2})a\sqrt{2}|<\Delta x/2
\end{equation}
will generate a comb state  $\ket{\bar 1}_{a\sqrt{2}}$.\\

\begin{figure}[!h]
\begin{center}
\includegraphics[width=8.6cm]{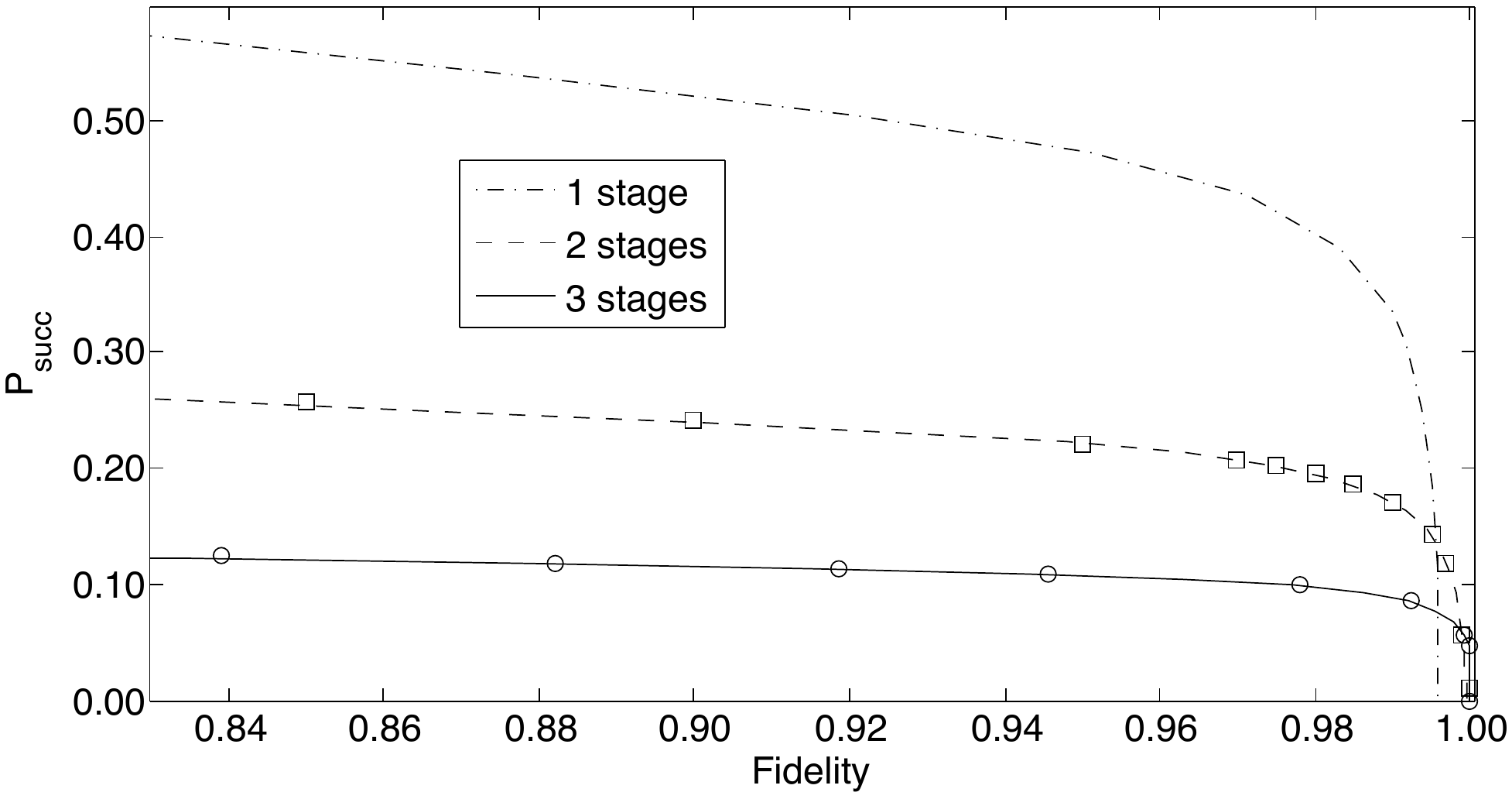}
\caption{{Same as Fig. \ref{fig2} for comb states generation with $p'$ iterations of the breeding stage of Fig. \ref{fig1} ($p'$ varying from $1$ to $3$) fed by a 5-stages breeded cat. Lines are obtained with $\Delta x_{l+1} = 1.3 \Delta x_l$, while the different symbols correspond to optimized $\Delta x_l$.}}
\label{fig3}
\end{center}
\end{figure}

 The numerical simulations, plotted in Fig. \ref{fig3} for up to $3$ iteration stages, were obtained using the heralding condition (\ref{HeraldingCond}). It appears that this breeding method allows to generate high fidelity comb states with a high efficiency. The empirical law $\Delta x_{l+1} = 1.3 \Delta x_l$ also applies in that case, and is very close to the best result.

Comb states may prove to have a wide range application in the field of quantum information. The purpose of the present paper is to show how they allow a loophole free violation of CHSH inequality with homodyne measurements, following the protocol proposed in \cite{Wenger}.

It is straightforward to check that the mixing of an even (resp. odd) squeezed cat state $\ket{\psi_+}^{s'}$ (resp. $\ket{\psi_-}^{s'}$) of amplitude $\alpha'$ with a comb state $\ket{\bar 1}_a^{s,s'}$ on a 50:50 beamsplitter creates a state $\ket{\mathfrak{f}}$ (resp. $\ket{\mathfrak{g}}$) which is very close to the state $\ket{f}$ (resp. $\ket{g}$) introduced in \cite{Wenger}:
\begin{eqnarray}
\braket{x}{\mathfrak{f}}\propto G_{1/s'}(x)\sum_k G_{\bar s}(x-(k+\frac{1}{2})\bar a)\cos[\frac{\pi x}{2\bar a}]&\qquad \\
\braket{x}{\mathfrak{g}}\propto G_{1/s'}(x)\sum_k G_{\bar s}(x-(k+\frac{1}{2})\bar a)\sin[\frac{\pi x}{2\bar a}]&\qquad
\label{fg_def}
\end{eqnarray}
with $\bar s=\sqrt{2}{s}$ and $\bar{a}=\sqrt{2}a$, under the condition $s'\alpha' a=\frac{\pi}{2\sqrt{2}}$.

The comb state can be obtained through $p'$ stages of comb breeding, fed by cats of amplitude $\alpha = \pi2^{p'/2}/(s'\bar a)$ and squeezing $s'$. Comparing the expressions of $\alpha$ and $\alpha'$ leads to $\alpha=\sqrt{2^{2+p'}}\alpha'$.
In other words, if the cats of amplitude $\alpha$ used for comb breeding are generated through $p$ cat breeding steps, the cat $\ket{\psi_+}^{s'}$ of amplitude $\alpha'$ has to be generated by $p-p'-2$ steps of cat breeding.

We are therefore able to perform the transformations $\ket{\psi_+}^{s'}\rightarrow\ket{\mathfrak{f}}$ and $\ket{\psi_-}^{s'}\rightarrow\ket{\mathfrak{g}}$. Starting from the two-mode state $\ket{\psi_+}^{s'}\ket{\psi_-}^{s'}+e^{i\theta}\ket{\psi_-}^{s'}\ket{\psi_+}^{s'}$, which can for instance be obtained through a delocalized photon subtraction \cite{Ourjoumtsev09}, we therefore generate $\ket{\mathfrak{f},\mathfrak{g}}+e^{i\theta}\ket{\mathfrak{g},\mathfrak{f}}$.
This last state is slightly different from the one proposed in \cite{Wenger}, but we can obtain
a violation of the CHSH inequality in the same way: quadrature $x$ or $p$ are measured, and one assigns $\epsilon_x=sign[\braket{x}{\mathfrak{f}}\braket{x}{\mathfrak{g}}]$ (if the measurement is $x$) or $\epsilon_p=sign[\braket{p}{\mathfrak{f}}\braket{p}{\mathfrak{g}}/i]$ (if the measurement is $p$) to the result. A maximal violation of the CHSH inequality $S\leqslant 2$ can then be obtained for $\theta\simeq-\pi/4$
(with $S:= |\!\!<\epsilon_p^A\epsilon_p^B>-<\epsilon_x^A\epsilon_x^B>\!\!|+|\!\!<\epsilon_p^A\epsilon_x^B>+<\epsilon_x^A\epsilon_p^B>\!\!|$).

\begin{figure}[!h]
\begin{center}
\includegraphics[width=8.6cm]{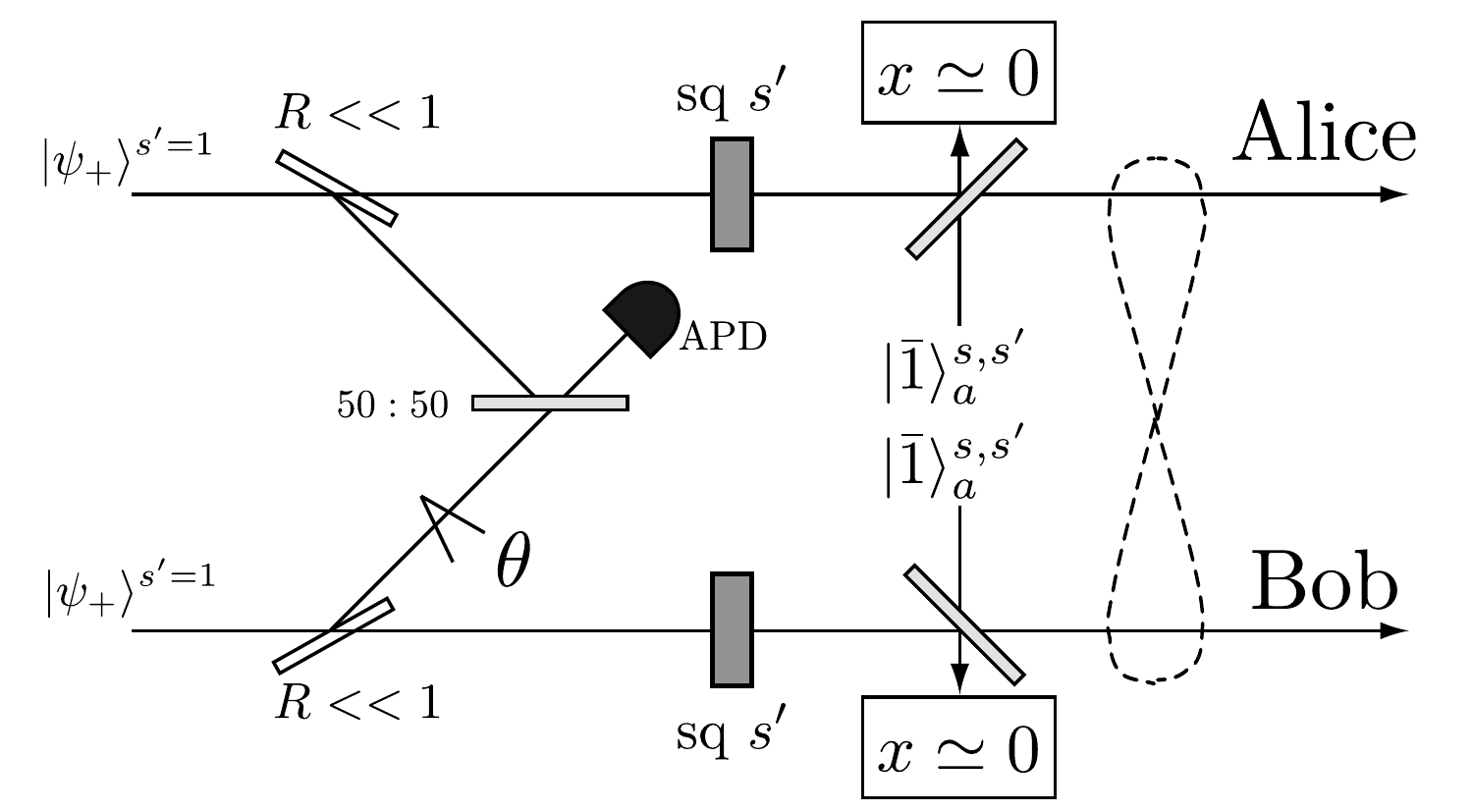}
\caption{{Proposal of a scheme for the generation of a state allowing a violation of the CHSH inequality with homodyne measurements.}} \label{fig4}
\end{center}
\end{figure}

Fig. \ref{fig4} pictures a possible implementation of this protocol: cat states $\ket{\psi_+}^{s'}$ have first to be unsqueezed ($\rightarrow s'=1$) before the delocalized photon subtraction, even if they have to be squeezed again (``sq" in Fig. 4) in order to match the comb state. After the photon subtraction, each mode is mixed with a comb state using the setup of Fig. \ref{fig1}, what we will call the ``modulation stage" and is based on a double heralding event. We are going to present results corresponding to two extreme cases: a ``strong" comb, obtained with $p=6$ stages of cat breeding and $p'=2$ stages of comb breeding, which will be called the state $(6,2)$; and a ``weak" comb, with an odd cat state obtained with $n=7$ photons instead of the comb $\ket{\bar{1}}$ ({\it i.e.} a comb state with $p'=0$), which will be called the state $(3,0)$.

The values obtained for the success probability of the whole setup depicted in Fig. 4 for different values of the heralding intervals $\Delta x$ as a function of $S$ is plotted in Fig. 5. $R$ has been taken equal to 0.001 and the total efficiency of the APD was taken equal to 0.06.

\begin{figure}[!h]
\begin{center}
\includegraphics[width=8.6cm]{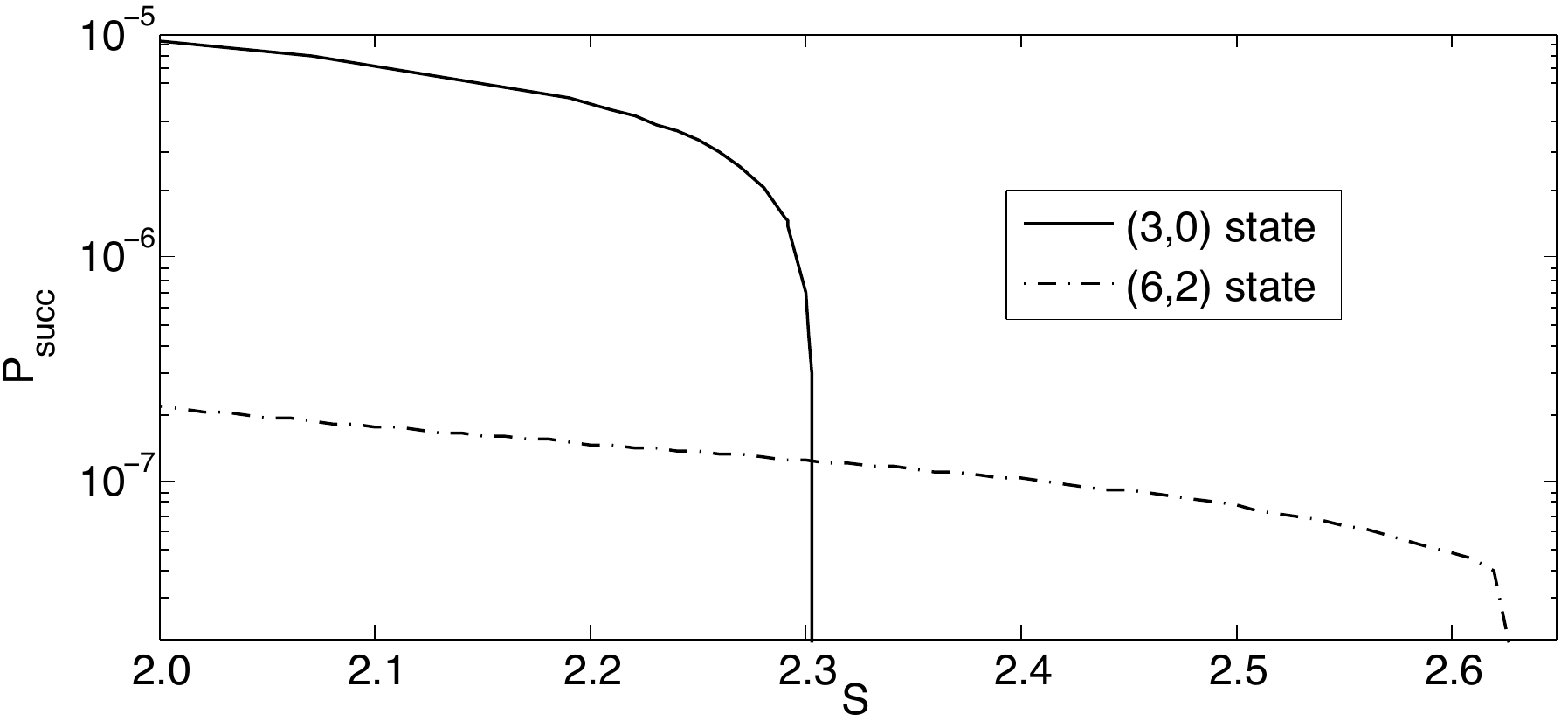}
\caption{{Total success probability of the protocol depicted in Fig. 4 for a comb state $\ket{\bar1}_a^{s,s'}=(3,0)$ and for $\ket{\bar1}_a^{s,s'}=(6,2)$ as a function of S.}} \label{fig5}
\end{center}
\end{figure}
Given the possibility to use high repetition rate sources in optics, this success probability could still lead to acceptable experiment durations.\\

\begin{figure}[!h]
\begin{center}
\includegraphics[width=8.6cm]{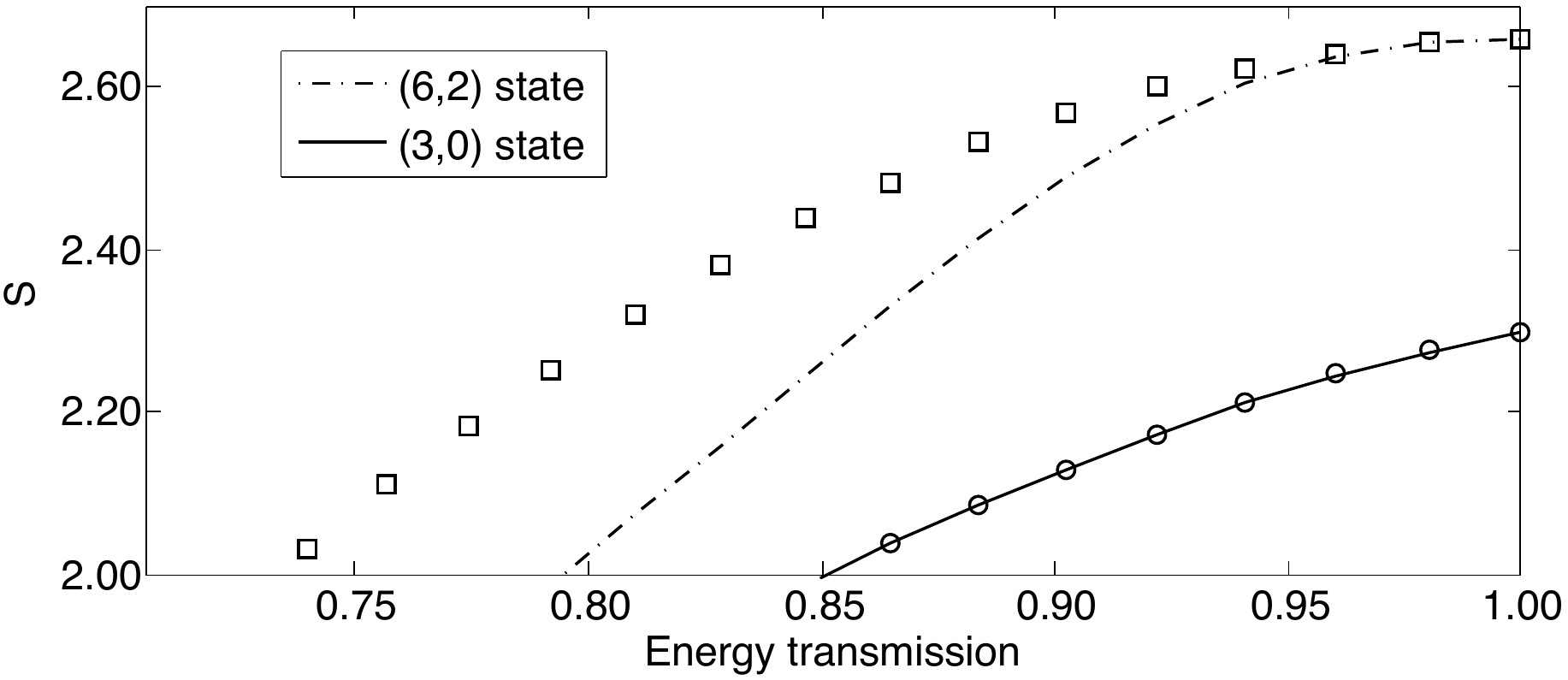}
\end{center}
\caption{{Lines: Value of $S$ obtained for comb states $(3,0)$ and $(6,2)$ as a function of line transmission; symbols: same as lines but with an additional squeezing in order to optimize robustness to losses.}} \label{fig6}
\end{figure}

A very impressive result concerns the robustness of these states against line losses. Fig. \ref{fig6} presents the evolution of $S$ with line transmission between the generated states and the final homodyne detections, and it can be seen that the (6,2) state permits a CHSH Bell inequality violation as long as $T > 74\%$ (energy transmission) if we adapt the squeezing of the state before undergoing losses. This figure also shows that the (6,2) state (10 photons in average) is more robust against losses than the (3,0) state (4 photons in average).

This shows that mesoscopic states can present interesting feature in term of robustness to losses, while still exhibiting strong quantum features.

\begin{acknowledgments}
We acknowledge support from the EU project ANR ERA-Net CHISTERA HIPERCOM.
\end{acknowledgments}

\bibliography{BibPaper0413}
\end{document}